\documentclass[aps,pra,reprint,twocolumn]{revtex4} 

\usepackage{graphicx}

\usepackage{hyperref} 
\usepackage{textcomp}
\usepackage{amstext}
\usepackage{amssymb,amsmath}

\newcommand{\ket}[1]{\left| #1 \right\rangle}

\begin{document}

\title{Analyzing quantum jumps of one and two atoms strongly coupled to an optical cavity}

\author{Sebastian Reick,$^{1,*}$ Klaus M{\o}lmer,$^2$ Wolfgang Alt,$^1$ Martin Eckstein,$^1$
Tobias Kampschulte,$^{1}$ Lingbo Kong,$^{1,\dag}$ Ren\'{e}
Reimann,$^1$ Alexander Thobe,$^{1,\ddag}$  Artur Widera,$^{1}$ and
Dieter Meschede$^{1}$}
\address{$^1$ Institut f\"{u}r Angewandte Physik der Universit\"{a}t Bonn, Wegelerstrasse 8, 53115 Bonn, Germany}
\address{$^2$ Lundbeck Foundation Theoretical Center for Quantum System Research,
Department of Physics and Astronomy, University of Aarhus, DK-8000
{\AA}rhus C, Denmark}
\address{$^{\dag}$ Present address: Department of Physics, East Carolina University, Greenville, NC,
USA}
\address{$^{\ddag}$Present address: Institut f\"{u}r Laser-Physik der
Universit\"{a}t Hamburg, Luruper Chaussee 149, 22761 Hamburg, Germany}
\address{$^*$Corresponding author: sreick@uni-bonn.de}

\begin{abstract}
We induce quantum jumps between the hyperfine ground states of one
and two Cesium atoms, strongly coupled to the mode of a high-finesse
optical resonator, and analyze the resulting random telegraph
signals. We identify experimental parameters to deduce the atomic
spin state non-destructively from the stream of photons transmitted
through the cavity, achieving a compromise between a good
signal-to-noise ratio and minimal measurement-induced perturbations.
In order to extract optimum information about the spin dynamics from
the photon count signal, a Bayesian update formalism is employed,
which yields time-dependent probabilities for the atoms to be in
either hyperfine state. We discuss the effect of super-Poissonian
photon number distributions caused by atomic motion.
\end{abstract}


\maketitle 

\section{Introduction}

Systems comprised of neutral atoms coupled to a single mode of a
high-finesse resonator belong to the key model-systems in quantum
optics \cite{Haroche06}. In the so called strong coupling limit an
atom periodically exchanges its excitation energy with the resonator
light field. In this case the dynamic evolution is governed by a
priori entangled light-matter quantum states, namely the combined
dressed states of the atom-cavity system. Due to the symmetric
interaction, described by the Janyes-Cummings Hamiltonian
\cite{Jaynes1963}, information about the state of the system can be
obtained from two complementary partial measurements: In the optical
domain, experiments rely on the detection of photons emitted from
the cavity \cite{McKeever2003, Boozer2006, Nussmann2005,
Nussmann2005a, Fortier2007, Khudaverdyan2008, Khudaverdyan2009},
whereas in the microwave regime the quantum state of atoms
transiting the cavity field is detected \cite{Raimond2001}.

Optical cavity-QED systems are attractive for applications in
quantum information science, e.g. for quantum networks. The
successful demonstration of, for instance, the mapping of the
coherent state of a traveling qubit (a photon) to the atomic state
memory qubit \cite{Boozer2007} as well as single photon generation
\cite{Wilk2007} are recent examples of significant progress in
controlling the interaction of a single atom with the cavity field.
For the creation of two-particle entangled states, promising
proposals rely on either applying deterministic protocols
\cite{You2003} or measurement induced (probabilistic) projection
\cite{Sorensen2003a, Metz2007}.

Not only for the application in quantum information science, but
also more generally for the investigation of strongly interacting
atom-cavity systems it is vital to understand the spin dynamics of
one and especially more than one atom simultaneously coupled to the
resonator field. In the work presented here we concentrate on the
case of one and two atoms and investigate how maximum information
about their hyperfine ground state can be retrieved from the stream
of photons arriving at the detector. We outline and detail the
identification of optimal experimental settings such as atom-cavity
detuning.

Random telegraph signals, obtained by continuously observing quantum
jumps between the spin states of a single atom, are analyzed by
applying a Bayesian update formalism. In complementary experiments
\cite{Gleyzes2007,Guerlin2007}, the photon number state (Fock state)
of a microwave cavity field is interrogated by a stream of circular
Rydberg-atoms acting as quantum probes. There Bayesian analysis has
proven to be a useful method of analysis, too. Random telegraph
fluctuations are a universal phenomenon observed in many different
fields, including a large variety of solid-state systems
\cite{Yuzhelevski2000}.

In order to study two-atom dynamics, we identified experimental
parameters for which the intra-cavity intensity depends on the
number of atoms in a specific spin state. The virtue of the Bayesian
method is evident in analyzing the corresponding telegraph signals,
for which the atomic state cannot be unambiguously deduced from the
measured transmission signal because of technical limitations on the
signal-to-noise ratio.

In our measurements we observe fluctuations in the transmission
exceeding shot noise, which we attribute to thermal motion of the
atom. We discuss the impact of this external dynamics on the
performance of the Bayesian analysis.

\section{Experimental techniques}

\subsection{Setup to trap and transport single atoms}

At the beginning of every experimental sequence, a controlled number
of cesium (Cs) atoms is transferred from a magneto-optical trap
(MOT) into a standing wave far-off-resonant dipole trap (FORT) with
$\lambda_{\text{FORT}}=1030$ nm and a trap depth of $U_{\text{FORT}}
\approx k_B \times 1$ mK. This trap acts as an ``optical conveyor
belt'' \cite{Kuhr2001} to transport atoms into the optical
resonator. The fundamental TEM$_{00}$ mode of the cavity has a
diameter of $2w_0 = 46$ \textmu m and a length given by the mirror
distance of 158 \textmu m, the finesse is $\mathcal{F} = 1.2 \times
10^6$.

\begin{figure}
\centering
\includegraphics[width=\columnwidth]{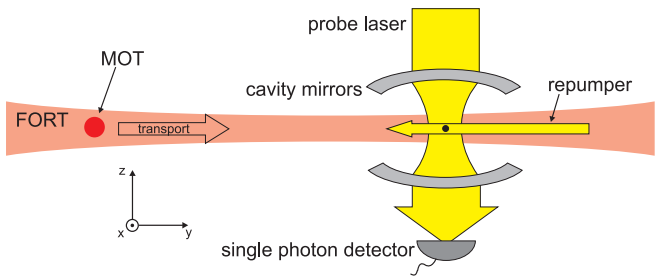}
\caption {Schematic setup of MOT, far-off resonance dipole trap
(FORT) and cavity mirrors (not to scale). Details on the
experimental setup and the stabilization of the cavity resonance
frequency are given in \cite{Khudaverdyan2008}.} \label{Reick-fig01}
\end{figure}

A conceptual drawing of the main components is depicted in figure
\ref{Reick-fig01}, for details on the cavity-setup and the
stabilization scheme see \cite{Khudaverdyan2008}. To study the
atom-cavity system, the transmission of a weak probe laser through
the cavity is detected with a single-photon counting module (SPCM).
Using a custom-build time-to-digital converter, we record - for each
photon click - the time since the last click, where for our typical
count-rates dead time effects are negligible. This list of
click-intervals is then converted into a binned transmission signal
by counting the detector clicks in each bin time interval $\Delta
t_b$.

The total detection efficiency for the probe laser light, including
absorption and scattering by the mirror coatings, losses at various
optical elements along the optical path, and the quantum efficiency
of the detector, amounts to $\eta = 4.4\%$, which is a threefold
improvement compared to our earlier work presented in
\cite{Khudaverdyan2009}. The main challenge was to optimize the
separation of probe- and stabilization-laser, with typical powers of
a few $10^{-15}$ and $10^{-6}$ Watts, respectively. In a first step,
they are separated by their carefully adjusted orthogonal
polarizations. Improved spectral filtering was achieved by replacing
a standard ruled diffraction grating with a volume holographic
grating, allowing us to omit an additional interference filter used
before, while still achieving a total suppression of the
stabilization laser to better than $10^{-8}$.

The probe laser frequency is set close to the \mbox{$|F=4\rangle
\rightarrow |F'=5\rangle$} transition of the Cs $D_2$ line, where
$F$ is the total angular momentum quantum number. For this
transition, the important parameters of the atom-cavity system are
$(g,\kappa,\gamma) = 2\pi\times(13.1,0.4,2.6)\;\text{MHz}$, where
$g$ is the nominal coupling strength for an atom at the position of
maximum coupling, $\kappa$ is the cavity field decay rate, and
$\gamma$ is the atomic dipole decay rate. Since in our setup the
birefringent splitting of the cavity resonances is larger than the
cavity linewidth, the cavity field is always linearly polarized,
causing a distribution of the atomic population over all Zeeman
sublevels due to photon scattering by the probe laser. Thus the
coupling strength $g$ given above is obtained from a weighted
average over all couplings $g(m_F)$, based on the steady state $m_F$
distribution for linearly polarized optical pumping
\cite{Gao1993,footnote1}. With the single-atom cooperativity
parameter $C_1 = g^2/(2\kappa\gamma) \gg 1$, our system is in the
strong coupling regime, where already a single atom significantly
influences the cavity spectrum.

\subsection{Nondestructive state detection}

\begin{figure}
\centering
\includegraphics[width=\columnwidth]{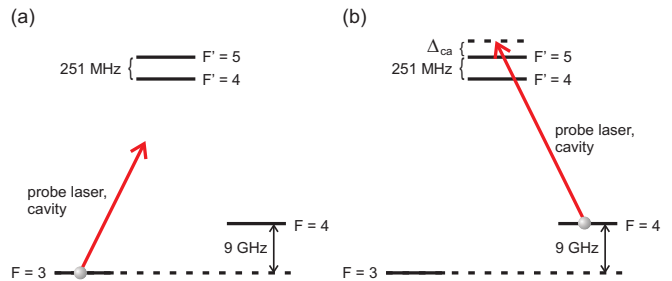}
\caption {Simplified Cs level scheme. (a) An atom in $\ket{F=3}$ is
so far detuned from the cavity resonance that it does not alter its
transmission. (b) If the atom is in $\ket{F=4}$, it changes the
transmission, depending on the cavity-atom detuning $\Delta_{ca}$
and the coupling strength $g$.} \label{Reick-fig02}
\end{figure}

In our system the two long-lived hyperfine groundstates $\ket{F=3}$
and $\ket{F=4}$ serve as qubit states \cite{Schrader2004}. For the
coupled atom-cavity system we measure this state by tuning the
cavity close to the $\ket{F=4}\rightarrow\ket{F'=5}$ transition,
where only an atom in the $\ket{F=4}$ state leads to a drop in the
transmission, while an atom in $\ket{F=3}$ is so far detuned (around
9.2 GHz) that it effectively decouples from the system and does not
influence the cavity transmission, see fig. \ref{Reick-fig02}.

The probe laser with angular frequency $\omega_p$ is initially tuned
to the resonance frequency of the empty cavity $\omega_c =
\omega_p$, so when an atom in $\ket{F=4}$ is inserted into the
cavity the transmission is reduced to a level which depends on the
detuning $\Delta_{ca} = \omega_c - \omega_a$, where $\omega_a$ is
the angular frequency of the atomic $\ket{F=4}\rightarrow\ket{F'=5}$
transition, including the AC-Stark shift induced by the FORT
potential. To experimentally distinguish between an atom in
$\ket{F=3}$ and an atom lost from the trap, which both result in the
same transmission signal, a repumping laser resonant with the
$\ket{F=3}\rightarrow\ket{F'=4}$ transition can be applied from the
side which brings the atom in $\ket{F=3}$ back to the
$\ket{F=4}\rightarrow\ket{F'=5}$ cycle. Thus for an empty cavity the
transmission would remain unchanged, while for an atom still present
in the cavity the transmission would drop again.

\section{Single atom spin dynamics}

If the state detection technique described above gave the same
result for an unlimited series of state measurements, it would be a
perfect projective quantum nondemolition (QND) measurement
\cite{Braginsky1980,Chaudhury2006,Windpassinger2008}, assuming the
system is otherwise unperturbed. However, in our situation the same
laser that we use to detect the atomic state (the probe laser) can
change it via inelastic hyperfine-state-changing Raman scattering.
An atom in the $\ket{F=4}$ groundstate can thus be transferred to
$\ket{F=3}$ via the $\ket{F'=3}$ and $\ket{F'=4}$ excited states,
thereby undergoing a \emph{quantum jump} \cite{Berquist1986,
Nagourney1986, Cook1985, Sauter1986,Itano1987}.

\subsection{Quantum jump rate and transmission level as a function of detuning}

\begin{figure}
\centering
\includegraphics[width=\columnwidth]{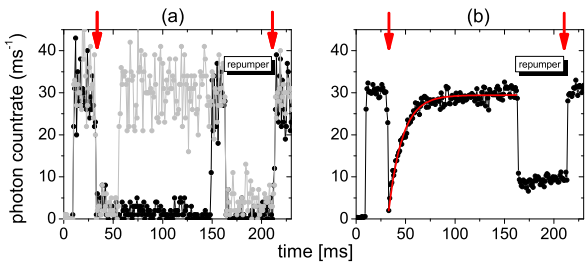}
\caption {(a): Black and grey curves show two single traces of
quantum jump measurements. The arrows indicate insertion and removal
of an atom. At the end of the sequence, the repumper is switched on
again to check that the atom was not lost. (b): Ensemble average
over 31 single traces. The average dwell time $R_{43}^{-1}$ is
obtained from the exponential fit. The averaged transmission level
at the end of the sequence, when the repumper is switched on, is
higher than the initial drop, indicating a lower average coupling
strength. This could be caused by increased thermal motion, a
re-distribution over different $m_F$-levels, or a combination of
both effects.} \label{Reick-fig03}
\end{figure}

To experimentally determine the rate $R_{43}$ of probe-laser-induced
transitions from $\ket{F=4}$ to $\ket{F=3}$ and to identify optimum
experimental conditions, we performed the following measurement: An
atom, optically pumped into $\ket{F=4}$, is transported into the
cavity center, causing a drop of the cavity transmission, see fig.
\ref{Reick-fig03} (a). Since no repumper is applied, probe-laser
scattering causes a spontaneous transition to $\ket{F=3}$ after some
time, visible as an instantaneous rise in transmission back to the
empty cavity level. To check whether the rise in transmission is
really due to a quantum jump and not caused by atom loss, the
repumper is switched on at the end of the sequence as discussed
above.

For each experimental realization, the quantum jump occurs at a
random point in time, see fig. \ref{Reick-fig03} (a) for two example
traces. Since the rate of state transitions is time independent, the
ensemble average plotted in fig. \ref{Reick-fig03} (b) reveals the
expected exponential curve with the time constant being the average
dwell time $R_{43}^{-1}$.

\begin{figure}
\centering
\includegraphics[width=\columnwidth]{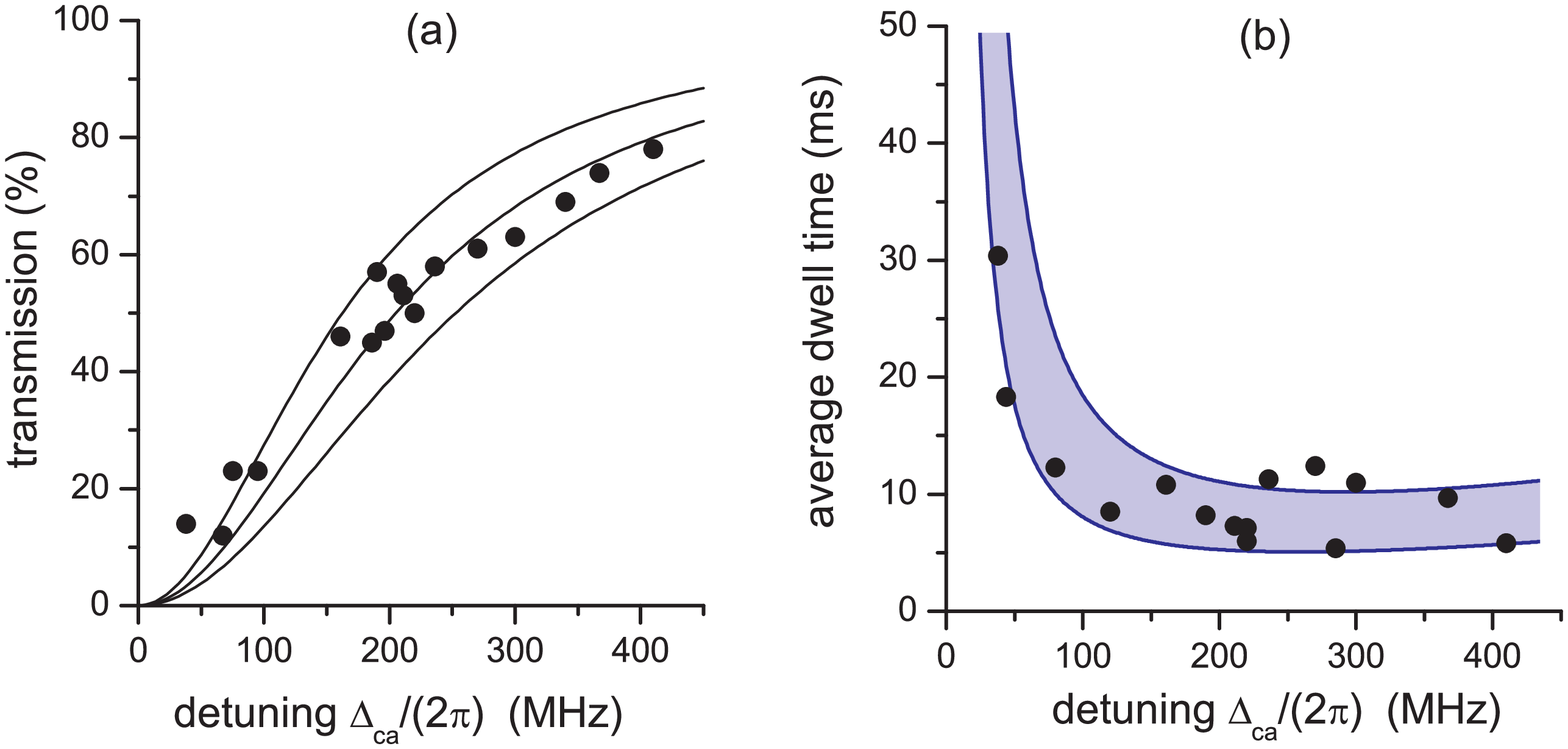}
\caption {(a) Normalized one-atom-transmission as a function of the
cavity-atom detuning $\Delta_{ca}$. The solid lines are calculated
for an atom at rest with an effective coupling strength of
$g_{\text{eff}}/(2\pi) = 8,9,$ and 10 MHz for the upper, middle, and
lower curve, respectively. (b) Average dwell time $R_{43}^{-1}$ as a
function of detuning. The shaded area is the result of a theoretical
model taking motion of the atom into account, and the range of
values represents our limited knowledge about the exact distribution
over the Zeeman-sublevels.} \label{Reick-fig04}
\end{figure}

This average dwell time was measured for a wide range of detunings
$\Delta_{ca}/(2\pi) = 38 ... 410 \;\text{MHz}$. For the same
settings, but with the repumper constantly applied, we measured the
transmission level $T_1$, defined as the photon count rate with one
atom in $\ket{F=4}$ coupled to the cavity, normalized to the empty
cavity signal. The results of both measurements are presented in
fig. \ref{Reick-fig04}.

In order to describe our measurements with a simplified analytical
model, we consider a two-level atom at rest with the probe-laser
being resonant with the empty cavity ($\omega_p = \omega_c$). In the
weak excitation regime, the one-atom-transmission level can be
expressed analytically as \cite{Hechenblaikner1998}
\begin{equation}
T_1(\Delta_{ca},g_{\text{eff}}) = \frac{\kappa^2 (\Delta_{ca}^2 +
\gamma^2)}{(\gamma \kappa + g_{\text{eff}}^2)^2 + (\Delta_{ca}
\kappa)^2} \;. \label{eq:TforZeroDeltaPC}
\end{equation}
The distribution over Zeeman sublevels, thermal motion of the atom,
AC-Stark shift variations, and other conceivable perturbations are
all accounted for by an effective coupling strength
$g_{\text{eff}}$. It is defined by Eq. (\ref{eq:TforZeroDeltaPC}) in
such a way that a stationary two-level atom with a coupling strength
of $g_{\text{eff}}$ would yield the experimentally measured
transmission level. The solid lines in fig. \ref{Reick-fig04} (a)
are calculated according to Eq. (\ref{eq:TforZeroDeltaPC}) with
$g_{\text{eff}}/(2\pi) = 8,9,$ and 10 MHz, and this range of
effective couplings describes the data reasonably well. We attribute
the difference between the nominal coupling strength of $g/(2\pi) =
$ 13.1 MHz and $g_{\text{eff}}$ mainly to thermal motion of the
atom.

To describe the measured average dwell times theoretically, see fig.
\ref{Reick-fig04} (b), $R_{43}$ is calculated as a function of
detuning  using the Kramers-Heisenberg-formula \cite{Cline1994}. For
this calculation, one has to treat the distribution over all Zeeman
sublevels and thermal motion separately, since this situation cannot
be modeled as a two-level system with an effective coupling. The
measured data agree satisfactorily with the theoretical model,
confirming that the best approximation to a projective
QND-measurement with longest dwell times is close to resonance. A
practical limitation is that stable coupling was never observed for
detunings $\Delta_{ca} \lesssim 2\pi\times$ 30 MHz, probably due to
cavity cooling becoming less effective \cite{Domokos2003,Murr2006}.

\subsection{Statistical analysis of single-atom random telegraph signals}\label{subsec:OneAtomBayes}

\begin{figure}
\centering
\includegraphics[width=\columnwidth]{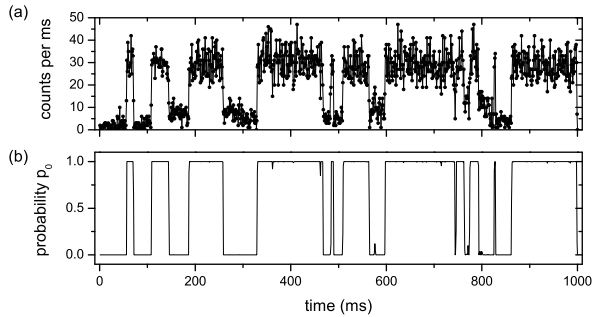}
\caption{(a) Random telegraph signal for one atom coupled to the
cavity. (b) Bayes analysis yielding $p_0(t)$, i.e. the probability
to be in $\ket{F=3}$.}\label{Reick-fig05}
\end{figure}

In the experiments discussed so far, the repumping laser was either
switched off or its intensity was adjusted such that an atom
off-resonantly transferred to $\ket{F=3}$ was pumped back to
$\ket{F=4}$ immediately, compared to all relevant time scales in our
experiment. In contrast, for the measurements presented in the
following, we deliberately attenuated the continuously applied
repumping laser to a level at which the transfer rate $R_{34}$ from
$\ket{F=3}$ to $\ket{F=4}$ was comparable to $R_{43}$. Therefore,
the resulting quantum jumps occur in both directions on a similar
timescale of several milliseconds and are thus detectable as a
random telegraph signal, see fig. \ref{Reick-fig05} (a) for an
example trace.

We quantify our knowledge about the atom's hyperfine spin by
probabilities assigned to the different atomic states. For the
following discussion we introduce the parameter $\alpha$ to denote
the number of atoms in $\ket{F=4}$. In this section $\alpha$ attains
only the values $0$ and $1$, whereas the case of two atoms (see
section \ref{sec:TwoAtoms}) also permits the value $\alpha=2$.
Internal state changes of a single atom are transitions between the
two states $\alpha=0,1$, and they occur with the rates $R_{10}$ and
$R_{01}$, which are identical to $R_{43}$ and $R_{34}$,
respectively. Although we imagine the transitions to occur randomly
and at discrete instances of time, the probabilities for the atom to
occupy the different states change in a continuous manner governed
by the following rate equations:
\begin{eqnarray}
\frac{\mathrm{d} p_0(t)}{\mathrm{d}t} &=&- R_{01} p_0(t) + R_{10} p_1(t)\;, \label{eq:SingleAtomREa}\\
\frac{\mathrm{d} p_1(t)}{\mathrm{d}t} &=&- R_{10} p_1(t) + R_{01}
p_0(t) = - \frac{\mathrm{d} p_0(t)}{\mathrm{d}t}\;.
\label{eq:SingleAtomREb}
\end{eqnarray}
The average steady-state probabilities $\bar{p}_0$ and $\bar{p}_1$
are obtained by setting $\frac{\mathrm{d} p_0(t)}{\mathrm{d}t} =
\frac{\mathrm{d} p_1(t)}{\mathrm{d}t} = 0$ and using $\bar{p}_0 +
\bar{p}_1 = 1$. The solutions are thus given by the ratio between
the transition rates:
\begin{eqnarray}
\bar{p}_0 = \frac{R_{10}}{R_{10} + R_{01}}\;, \label{eq:steadyStateProbA}\\
\bar{p}_1 = \frac{R_{01}}{R_{10} +
R_{01}}\;.\label{eq:steadyStateProbB}
\end{eqnarray}
The average probabilities and thus the ratio of the rates $R_{10}$
and $R_{01}$ can therefore be obtained from photon count histograms
by the following procedure: Along with the telegraph signals,
transmission traces for an empty cavity ($\alpha = 0$), and for one
continuously coupled atom ($\alpha=1$) were measured for otherwise
identical settings. From these three sets of data, normalized photon
count histograms $\mathcal{P}(n)$, $\mathcal{P}(n|0)$ and
$\mathcal{P}(n|1)$ are computed, with $n$ being the number of
photons detected per binning time $\Delta t_b = 1$ ms. Here and for
the remaining discussion, $\mathcal{P}$ always refers to
photon-count probabilities, while $p_{\alpha}(t)$ indicates
spin-state probabilities.

Since the telegraph signal is expected to represent the atomic
system jumping between the different states, the associated
accumulated histogram of photon counts should be a weighted fit
\begin{equation}
\mathcal{P}(n) = \bar{p}_0 \mathcal{P}(n|0) + (1-\bar{p}_0)
\mathcal{P}(n|1),
\end{equation}
of the independently measured histograms $\mathcal{P}(n|\alpha)$ for
the two atomic states. Treating $ \bar{p}_0$ as a fitting parameter
yields $\bar{p}_0 = 0.64$ and $\bar{p}_1 = 0.36$.

\begin{figure}
\centering
\includegraphics[width=\columnwidth]{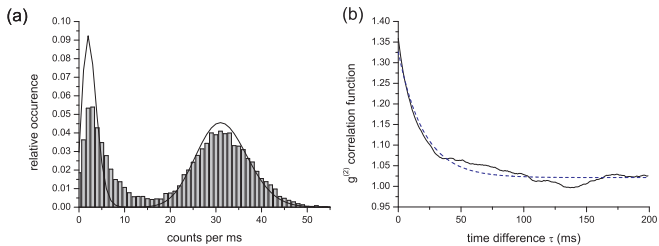}
\caption {(a) Normalized histogram extracted from 13 telegraph
signals of 1000 ms duration each, binned with $\Delta t_b = 1$ ms.
The solid line is a Poissonian distribution with the same maximum
count rate. (b) Averaged second order correlation function $g^{(2)}$
for the same set of telegraph signals. The blue dashed line is an
exponential fit yielding the time constant $(R_{10}+R_{01})^{-1} =
20$ ms.} \label{Reick-fig06}
\end{figure}

In order to extract the transition rates, we note that the jumping
of the atom between two different states with different transmission
properties causes characteristic fluctuations in the number of
detection events obtained in different time bins, $n(t)$ and
$n(t+\tau)$, which become visible in the second-order correlation
function $g^{(2)}(\tau)$. Assuming Poissonian count distributions,
an analysis of the rate equations yields \cite{Cook1985}
\begin{equation}
g^{(2)}(\tau) = \frac{\langle n(t)n(t+\tau)\rangle}{\langle
n(t)\rangle \langle n(t+\tau)\rangle} \propto
\exp{(-(R_{10}+R_{01})\tau)} \;\text{for}\; \tau > 0\;.
\;\label{eq:g2rates}
\end{equation}
The histogram of the telegraph signal and the $g^{(2)}$ function are
plotted in figure \ref{Reick-fig06} (a) and (b), respectively. From
an exponential fit of the correlation function, we get
$R_{10}+R_{01} =$ \mbox{50 s$^{-1}$}, therefore we obtain $R_{10} =$
\mbox{40 s$^{-1}$} and $R_{01} =$ \mbox{18 s$^{-1}$} using
$\bar{p}_0$, $\bar{p}_1$, and Eqs. (\ref{eq:steadyStateProbA}) and
(\ref{eq:steadyStateProbB}).

In the discussion above, we assumed that the state of the atom can
be described by the two states $\alpha=0$ and $\alpha=1$ alone, each
leading to a Poissonian distribution $\mathcal{P}(n|\alpha)$ of the
photon count rate . For $\alpha=0$ this is verified by the
measurement: The right peak of the measured histogram in figure
\ref{Reick-fig06} (a) agrees with a Poissonian distribution of the
the same average count rate. Thus the state detection for
$\alpha=0$, with the transmission being equal to the empty-cavity
case, is essentially shot-noise limited and residual frequency or
intensity fluctuations of the probe-laser can be neglected.

However, comparing the photon count histogram $\mathcal{P}(n|1)$
with the Poissonian distribution (left peak of the histogram in fig.
\ref{Reick-fig06} (a))  indicates super-Poissonian fluctuations. We
attribute these mainly to thermal motion of the atom: The coupling
constant $g$ follows the cavity mode function, i.e.
\mbox{$g(\mathbf{r}) = g_0 \psi(\mathbf{r})$}, which in turn leads
to a transmission level $T_1(\mathbf{r})$, depending on the atomic
position, according to Eq. (\ref{eq:TforZeroDeltaPC}).

In later sections of this paper we shall discuss candidates for a
more complete theoretical analysis of this dynamics. At this point,
we pursue a pragmatic approach and still extract the atomic
transition rates from the correlation function as stated by Eq.
(\ref{eq:g2rates}), because this relation does not rely strongly on
the Poissonian character of the signal. Furthermore, the exact
values of the rates are not the main result of this work and do not
convey fundamental physical insight, since they are determined by
the intensities of the probe and repumping laser. They rather
constitute parameters in the following statistical analysis.

To quantify the knowledge about the atomic state that we obtain from
the measured telegraph signals, we use a Bayesian approach in
analyzing the data. The philosophy behind this approach is that we
assign probabilities to the possible states $\alpha=0,1$ of the
atom, and acknowledge that these probabilities merely reflect our
incomplete knowledge about the system, unless one of the
probabilities is unity. Due to the atomic transitions which occur
without our direct noticing, the probabilities of the unobserved
system obey the rate equations (\ref{eq:SingleAtomREa}) and
(\ref{eq:SingleAtomREb}), but since  the cavity transmission depends
on the atomic state, we learn about the atomic state from the
observed photon count record.

The probabilities $p_{\alpha}$ are thus calculated step-wise from
the incremental information obtained in every time bin of the
measured telegraph signal. Let $n(t_i)$ be the number of photons
detected during the interval $[t_i - \Delta t_b/2, t_i+\Delta
t_b/2]$, where the binning time $\Delta t_b$ is fixed to 1\,ms for
the following analysis. With $p_{\alpha}(t_i)$ we refer to the
probability for an atom to be in the state $\alpha$ in the midpoint
of the aforementioned interval. Assuming that the atomic state
probabilities in the previous time bin $p_{\alpha}(t_{i-1})$ are
known, the probabilities $p_{\alpha}(t_i)$ are estimated by first
evolving their values according to the rate equations
(\ref{eq:SingleAtomREa}) and (\ref{eq:SingleAtomREb}). In a linear
approximation for $R_x \Delta t_b \ll 1$, where $R_x =
\text{max}(R_{10},R_{01})$, this leads to
\begin{eqnarray}
\tilde{p}_0(t_i)  &=& p_0(t_{i-1}) + (R_{10} \: p_1(t_{i-1}) -R_{01}\: p_0(t_{i-1}) ) \Delta t_b \;,\\
\tilde{p}_1(t_i)  &=& p_1(t_{i-1}) + (R_{01} \: p_0(t_{i-1}) -R_{10}
\:p_1(t_{i-1})) \Delta t_b \;,
\end{eqnarray}
where $\tilde{p}$ indicates the unconditional probability.

Note that the probabilistic description does not imply that the atom
occupies two different states, but only that we do not know which
one is actually occupied. This also implies that our prediction for
the distribution of photon numbers $n(t_i)$ detected in the $i$-th
time bin has to be calculated as a weighted average $\mathcal{P}(n)
= \tilde{p}_0 \mathcal{P}(n|0) + (1-\tilde{p}_0) \mathcal{P}(n|1)$.
The actually measured photon counts $n(t_i)$ provide new
information, and the state probabilities are updated using Bayes'
rule of conditional probabilities:
\begin{equation}
p_{\alpha}(t_i) \equiv p(\alpha|n(t_i)) =
\frac{\mathcal{P}(n(t_i)|\alpha)\tilde{p}_{\alpha}(t_i)}{\sum_{\alpha}
\tilde{p}_{\alpha}(t_i)
\mathcal{P}(n(t_i)|\alpha)}\;\text{for}\;\alpha =
0\;\text{and}\;1\;.
\end{equation}
The conditional probabilities $\mathcal{P}(n(t_i)|\alpha)$ are
extracted from the separately measured photon count histograms for
$\alpha=$ 0 and 1, respectively. Setting the initial probabilities
to $p_0(0) = 0$, $p_1(0) = 1$, because the atom is prepared in
$\ket{F=4}$ before being transported into the cavity,
$p_{\alpha}(t_i)$ is then updated stepwise for each time bin. In
this way, the time-dependent atomic state probabilities are computed
successively for the whole transmission trace.

Figure \ref{Reick-fig05} (a) shows an example trace of a telegraph
signal to which the Bayesian algorithm was applied. Most of the
time, the probability $p_0(t)$, plotted in (b), is close to either 0
or 1, while narrow spikes indicate short periods of time with less
complete knowledge about $p_{\alpha}$. The Bayes analysis provides
more definite probabilities than a matching of the currently
transmitted signal to the state dependent transmission rate, because
it updates previously estimated results and thus accumulates
statistical significance over time. The optical probing of the
system does not, however, prevent atomic transitions from taking
place, and during such transitions, the Bayes algorithm faithfully
reproduces our inability to determine the state of the atom with
certainty until a significant amount of data has been accumulated
which is in agreement with the new state of the atom. In this
context, the narrow spikes in the figure illustrate the
``willingness" of the Bayesian update to interpret a few unexpected
photon counts as the emerging signal of a change of state, while
they may be only statistical fluctuations. For photon count
histograms with negligible overlap, the Bayesian algorithm would
yield the same result as a simple threshold analysis. Its main
advantage is that one can still extract information about the spin
dynamics even for a signal where the signal-to-noise ratio prohibits
a threshold analysis, as will become apparent in the next section.

\section{Spin dynamics of two atoms}\label{sec:TwoAtoms}

So far we have presented experiments revealing the internal spin
dynamics of one atom coupled to the cavity mode. Placing two atoms
into the resonator leads to an effective interaction between them,
mediated by the cavity field \cite{You2003}, and detecting the
number of atoms being in a particular state could be used for
entanglement generation in cavity-QED-systems \cite{Sorensen2003a}.

\subsection{Counting the number of atoms in $\ket{F=4}$}

In the previous section the atomic state was determined from the
probe laser transmission. Without changing the experimental
settings, this is not directly possible for two atoms coupled to the
resonator. Both atoms in $\ket{F=3}$, i.e. $\alpha=0$, will lead to
a transmission level $T_0=1$ equal to the empty-cavity case. One
atom in $\ket{F=4}$ and one in $\ket{F=3}$ ($\alpha=1$) will cause
the transmission $T_1$ to drop almost to zero, which implies that
$\alpha=2$ is indistinguishable from $\alpha=1$. To deduce
$\alpha=0,1,2$ from the corresponding transmission levels $T_0, T_1,
T_2$, the experimental settings have to be adapted.

\begin{figure}
\centering
\includegraphics[width=\columnwidth]{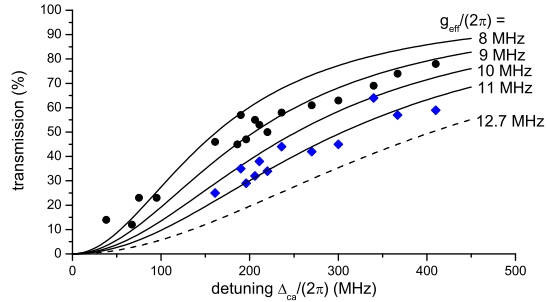}
\caption {Normalized transmission $T_1$ (black dots) and $T_2$ (blue
diamonds) for one and two atoms, respectively. The solid lines are
calculated according to the effective two-level model
(\ref{eq:TforZeroDeltaPC}) for one atom at rest with different
values for $g_{\text{eff}}$, and the dashed line shows the
theoretically expected two-atom transmission for
$g_{\text{eff}}/(2\pi)=\sqrt{2}\times9$ = 12.7 MHz. The one-atom
data is the same as in fig. \ref{Reick-fig04} (a).}
\label{Reick-fig07}
\end{figure}

In the weak excitation limit, two atoms at rest coupled with the
same strength $g$ to the cavity can be theoretically described as a
single atom experiencing a coupling strength $g_2 = \sqrt{2} g$. In
the dispersive limit ($\Delta_{ca} \gg \gamma$), Eq.
(\ref{eq:TforZeroDeltaPC}) thus yields
\begin{equation}
T_1 = \frac{1}{1 + \left(\frac{g^2}{\kappa\Delta_{ca}}\right)^2}
\;,\; T_2 = \frac{1}{1 + \left(\frac{2
g^2}{\kappa\Delta_{ca}}\right)^2} \;, \label{eq:T1T2}
\end{equation}
for the transmission levels. The level difference $\Delta T_{12} =
T_1 - T_2$ reaches its maximum value of 33\% for $g^2/(\kappa
\Delta_{ca}) = 1/\sqrt{2}$, where $T_0$, $T_1$, and $T_2$ are
equally spaced. In order to examine this theoretical prediction
experimentally, the transmission level $T_2$ was measured alongside
the one-atom transmission. Figure \ref{Reick-fig07} shows that for
two atoms the transmission is lower, but instead of the
theoretically expected value of $\sqrt{2}\times9\,$MHz
$\approx13\,$MHz, it is compatible with an effective coupling of
$g_{\text{2,eff}} \approx 2\pi\times 11$ MHz. As a consequence, the
measured level difference $\Delta T_{12}$ is at maximum about 20\%
for a detuning of $\Delta_{ca} = 2\pi\times270$ MHz.

A detuning of $\Delta_{ca}$ in the range of 200 to 300 MHz has,
however, two disadvantages for studying the spin-dynamics of two
coupled atoms: Firstly, the difference in cavity transmission is
quite small compared to the noise, and secondly, the average dwell
time $R_{43}^{-1}$ is close to its minimum value for $\Delta_{ca} >
2\pi\times 150$ MHz, with a very shallow slope towards higher
detunings, see fig. \ref{Reick-fig04} (b). Closer to resonance, this
time is longer, but if two atoms are at the cavity center, the
transmission levels $T_1$ and $T_2$ are almost indistinguishable.

\begin{figure}
\centering
\includegraphics[width=\columnwidth]{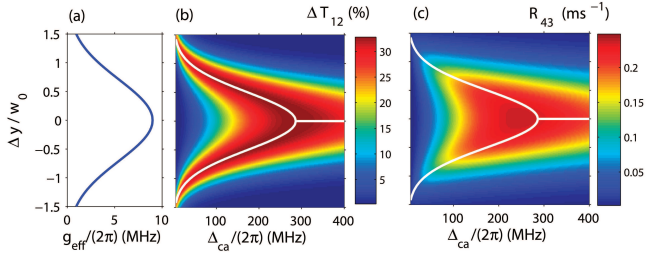}
\caption {(a) Effective coupling as a function of distance $\Delta
y$ from the cavity center. (b) Transmission level difference $\Delta
T_{12}$, (c) quantum jump rate $R_{43}$ as a function of detuning
$\Delta_{ca}$ and distance $\Delta y$. The white solid lines are
points of maximum $\Delta T_{12}$ calculated according to Eq.
(\ref{eq:DeltaY}). } \label{Reick-fig08}
\end{figure}

The level difference $\Delta T_{12}$ can, however, be controlled for
a constant detuning $\Delta_{ca}$ by changing $g_{\text{eff}}$. This
is possible by means of our optical conveyor belt, which allows us
to transport atoms not only into the cavity center, but also to stop
the transport at a predetermined distance $\Delta y$ away from it.
With $g_{\text{eff}}(\Delta y=0) = 2\pi\times 9$ MHz, the coupling
strength as a function of position along the conveyor belt axis
reads $g_{\text{eff}}(\Delta y) = g_{\text{eff}}(0) \exp(-\Delta
y^2/w_0^2)$. From Eq. (\ref{eq:T1T2}) the required distance $\Delta
y$ to achieve $\Delta T_{12} = 0.33$ is calculated to be
\begin{equation} |\Delta y(\Delta_{ca})| = w_0 \sqrt{\frac{1}{2}
\ln\left(\frac{\sqrt{2} g_{\text{eff}}^2(0)}{\Delta_{ca}
\kappa}\right)}\;.\label{eq:DeltaY}
\end{equation}
For $\Delta_{ca} > 2\pi\times 280$ MHz, $\Delta T_{12}$ is always at
maximum for $\Delta y = 0$, i.e. at the cavity center. Figure
\ref{Reick-fig08} shows the calculated level difference $\Delta
T_{12}$ and the quantum jump rate $R_{43}$ as a function of detuning
$\Delta_{ca}$ and distance from the cavity center $\Delta y$. By
choosing a lower detuning, the scattering rate $R_{43}$ is reduced,
and it is still possible to obtain optimal distinction $\Delta
T_{12}$ by positioning the atoms away from the cavity center.
Empirically we found that a detuning of $\Delta_{ca} = 2\pi\times
38$ MHz  is a lower limit in terms of stable transmission traces.
The distance of $\Delta y = 21$ \textmu m, at which
$g_{\text{eff}}/(2\pi) \approx 3.1$ MHz, was adjusted for optimum
distinction of one and two atoms.

\subsection{Two-atom telegraph signal}

To study two-atom spin dynamics, two atoms loaded into the FORT were
positioned at $\Delta y = 21$ \textmu m. At this position of around
one cavity-waist away from the mode center, the coupling strength
depends more critical on the exact position, therefore those traces
were selected for which the measured atom-atom spacing was
$\leq2$\,\textmu m.  As for the one-atom case, the repumper was
attenuated to a level at which it induced quantum jumps from
$\ket{F=3}$ to $\ket{F=4}$ at a rate comparable with the probe laser
induced jumps.

\begin{figure}
\centering
\includegraphics[width=\columnwidth]{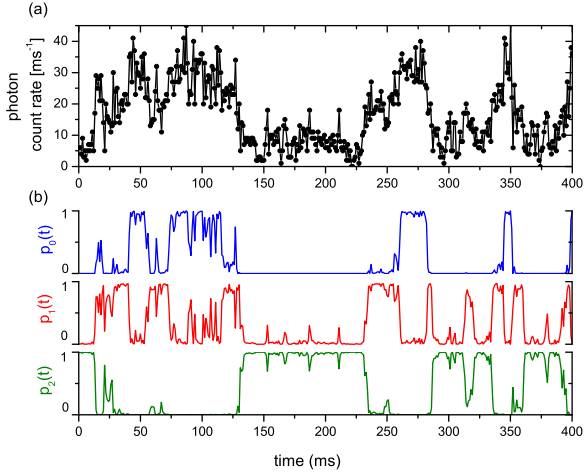}
\caption {(a) Example trace of a random telegraph signal for two
atoms placed $\Delta y=21$ \textmu m away from the cavity center.
The cavity-atom detuning is $\Delta_{ca} = 2\pi\times 38$ MHz. (b)
Probabilities for 0,1, or 2 atoms to be in $\ket{F=4}$, calculated
using the Bayes method.}\label{Reick-fig09}
\end{figure}

\begin{figure}
\centering
\includegraphics[width=\columnwidth]{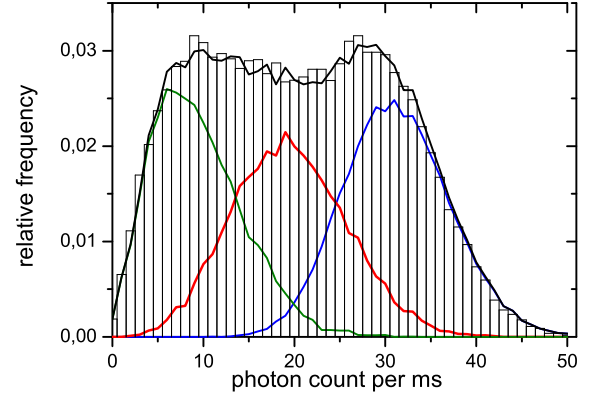}
\caption {Normalized photon count histogram (bars) of many two-atom
telegraph signals. The blue, red, and green lines are independently
measured histograms for 0, 1, and 2 atoms coupled continuously to
the cavity, respectively. The black line is a weighted sum of those
three histograms.}\label{Reick-fig10}.
\end{figure}

Figure \ref{Reick-fig09} (a) shows an example single trace of a
two-atom telegraph signal. For \mbox{$t \approx 200 \ldots
300$\,ms}, steps corresponding to $\alpha = 2$ (low transmission),
$\alpha=1$ (intermediate level), and $\alpha=0$ (empty cavity
transmission) are discernible, but in general the distinction
between the levels is not as clear as for the one-atom case. The
degree of the level separation can be deduced from a histogram
extracted from several hundred telegraph signals, see fig.
\ref{Reick-fig10}.

This histogram does obviously not show a three-peak structure. To
quantify the contributions of the transmission levels $T_0$, $T_1$,
and $T_2$, we independently measured photon count histograms for 0,
1, and 2 atoms coupled to the resonator at the same position and for
the same detuning as for the telegraph signals, depicted as solid
lines in fig. \ref{Reick-fig10}. These were obtained from signals of
continuously coupled atoms, i.e. a sufficiently strong repumper was
applied. The photon count histogram of the telegraph signal (black
line) agrees well with a fit calculated as a weighted sum of the
three individual histograms conditioned on the atomic states.

The statistical analysis is performed analogous to the one atom
case, but the set of rate equations now involves three atomic states
and reads
\begin{eqnarray}
\frac{d p_0}{dt} &=& -  R_{01} p_0(t) + R_{10} p_1 (t) \;,\\
\frac{d p_1}{dt} &=&  R_{01} p_0(t)  - R_{10} p_1 (t) - R_{12} p_1(t) + R_{21} p_2(t)\;, \\
\frac{d p_2}{dt} &=&  R_{12} p_1(t) - R_{21} p_2(t) \;.
\end{eqnarray}
A transition of an atom from $\ket{F=3}$ to $\ket{F=4}$ is only
induced by the repumper at a rate $R_{\text{rep}}$, which is
independent of $\alpha$ because the laser is applied from the side
of the cavity. Thus $R_{12} = R_{\text{rep}}$ and $R_{01} = 2
R_{\text{rep}}$, because for the latter case two atoms both in
$\ket{F=3}$ are present. In contrast, $R_{21}$, i.e. the rate that
one out of two atoms in $\ket{F=4}$ undergoes a quantum jump to
$\ket{F=3}$, is \emph{not} simply given by $2 R_{10}$, because this
transition is induced by the probe laser, the intensity of which
depends on $\alpha$ \cite{Khudaverdyan2009}. Theoretically, the jump
rate depends linearly on the intracavity intensity, thus we expect
\begin{equation}
R_{21} = 2 \frac{T_2}{T_1} R_{10} \;,\label{eq:rateRatio}
\end{equation}
but this relation is not fixed for the calculation and the three
rates $R_{21}$, $R_{21}$ and $R_{\text{rep}}$ are considered as
independent parameters for the calculation. A weighted fit to the
photon count histogram has two independent fit parameters and yields
the steady state populations, which are related to the ratio of the
three rates. In contrast to the single atom case, here it is not
possible to the make a reasonable fit to the correlation function to
obtain the sum of the rates and thus all three parameters. Instead,
we initially guess the transition rates and employ the Bayesian
update method to extract time dependent atomic state probabilities.
Then we apply a fit as described below to iteratively extract values
for the transition rates $R_{10}$, $R_{21}$ and $R_{\text{rep}}$,
which ensure the optimum agreement of the time averaged
probabilities with the steady state solution of the rate equations.

A good initial guess for the rate $R_{10}$ can be obtained from the
transition rate for a \emph{single} atom placed at the same distance
$\Delta y$ away from the cavity center, with no repumper applied,
similar to the measurement presented in fig. \ref{Reick-fig03} (b).
The transmission levels $T_2$ and $T_1$ are measured independently,
which yields then an estimate for $R_{21}$ according to Eq.
(\ref{eq:rateRatio}). The rate $R_{\text{rep}}$ cannot be measured
independently, but since the power of the repumping laser is
adjusted such that the transition rates from $\ket{F=4}$ to
$\ket{F=3}$ and vice versa are approximately equal, $R_{\text{rep}}$
is set to $R_{10}$ as a starting value for the calculation.

With the initial probabilities $p_0(0) = 0, p_1(0) = 0, p_2(0) =1$,
the Bayesian algorithm is performed step-wise for each time bin as
described for the one-atom case, yielding probabilities
$p_{\alpha}(t)$. Improved values of the three transition rates are
now determined by the following iterative, self-consistent method:

An analytical solution of the rate equations for $p_0(t)$, $p_1(t)$
and $p_2(t)$, with the initial conditions given above, yields the
ensemble-averaged probabilities $\langle p_{\alpha} \rangle (t)$
with the three jump rates as parameters. Averaging over the
probabilities $p_{\alpha}(t)$ obtained from the analysis of many
traces provides an \emph{experimental} result for $\langle
p_{\alpha} \rangle (t)$, which can be fitted with the analytical
solution, in which the rates $R_{10}$, $R_{21}$ and $R_{\text{rep}}$
are used as fit parameters. With the new values for the rates
obtained in this way, the Bayes algorithm is applied over again to
all experimental traces, yielding an updated set of time dependent
probabilities $p_{\alpha}(t)$, which is again averaged to extract
the rates, etc. The converged set of rates obtained from this
analysis is
\begin{equation}
R_{10} = 104\,\mathrm{s}^{-1}\;,\; R_{21} =
52\,\mathrm{s}^{-1}\;,\;R_{\text{rep}} = 45\,\mathrm{s}^{-1}\;,
\end{equation}
and the final results for $p_{\alpha}(t)$ for the example trace are
shown in fig. Fig. \ref{Reick-fig09} (b). The ratio between $R_{10}$
and $R_{21}$ obtained from this iterative process does not confirm
the assumption of Eq. (\ref{eq:rateRatio}), because with $T_1
\approx 2 T_2$, we would expect $R_{10} \approx R_{21}$. The reason
for this discrepancy remains unclear at this stage.

\section{Discussion of statistical analysis}

In this section we will address some questions arising in connection
with the statistical analysis presented in the manuscript. First, we
will discuss the dependence of the Bayesian atomic state analysis on
the measurement data binning time, which presents interesting
questions both in the case of Posissonian and non-Poissonian
counting statistics. Secondly, we will discuss the possible origin
of the non-Poissonian character of the photon count records and its
consequences for our extraction of rate parameters and the Bayesian
analysis.

\subsection{Bin size and optimum information extraction}

\begin{figure}
\centering
\includegraphics[width=\columnwidth]{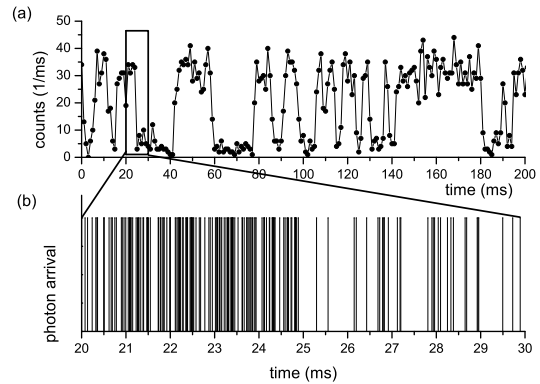}
\caption{(a) Random telegraph signal with 1 ms binning time. (b)
Enlarged section of 10 ms showing photon click times. The quantum
jump occurs at about 24.8 ms. }\label{Reick-fig11}
\end{figure}

In the analysis of the one and two atom telegraph signals discussed
so far, we used binning times of 1 ms. Let us recall that the
raw-data of the cavity transmission consists of a list of time
intervals between photon clicks, see Fig. \ref{Reick-fig11} (b) for
an example trace. To study some of the consequences which a change
in time bin size might have, we analyzed one and the same set of
data using the Bayes formalism, but for different bin sizes.

If long time bins are used, the signal to noise ratio in each bin is
good, and the count histograms for each atomic state become well
separated. This implies that for long sequences of time, the atomic
state probabilities will be firmly fixed to values close to zero and
unity, while the instances where transitions between the states
occur are not resolved within the duration of a single time bin. But
this is only true as long as $\Delta t_b < R^{-1}$, with
$R=\text{max}(R_{01},R_{10})$, because for even longer times
transitions will occur within a significant fraction of the bins
causing considerable uncertainty about the actual atomic state.

\begin{figure*}
\centering
\includegraphics[width=0.7\textwidth]{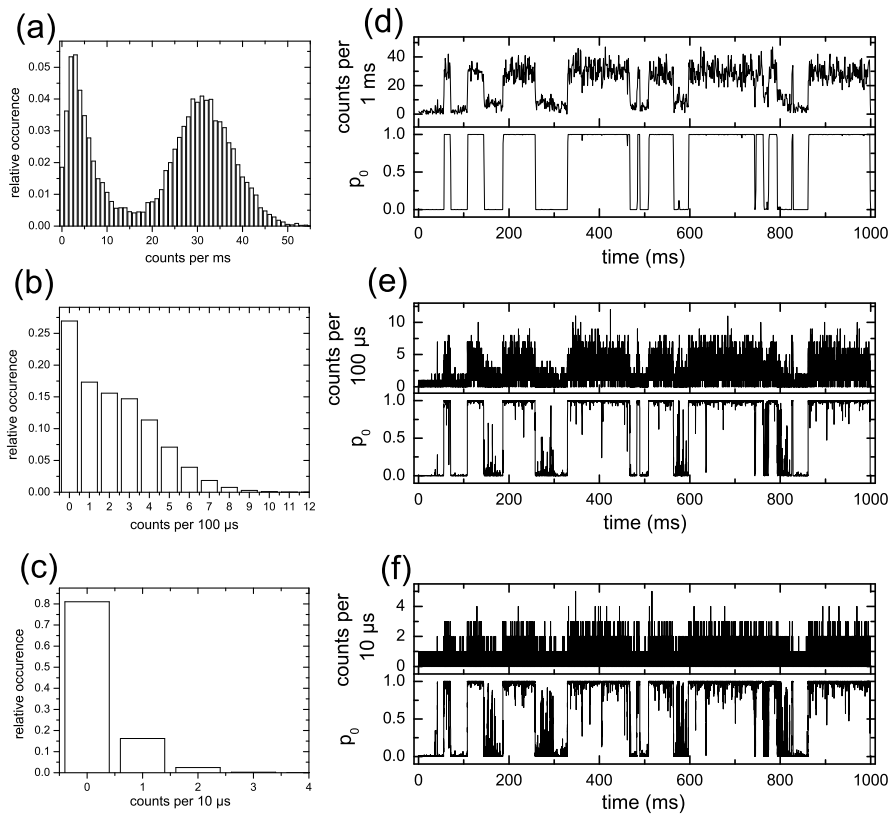}
\caption{Application of the Bayes algorithm for different bin times.
Figures (a) to (c) show the histogram of the telegraph signal for a
bin time of 1 ms, 100 \textmu s, and 10 \textmu s, respectively. The
transmission signals, generated from the same photon-click record,
and $p_0(t)$ are depicted in (d) to (e).}\label{Reick-fig12}
\end{figure*}

Going to shorter time bins, the signal-to-noise ratio is decreased,
and the overlap of the photon count histograms become larger.
Correspondingly, it happens more frequently that a less probable,
but still possible, number of counts in a time bin causes a narrow
spike in the atomic state probabilities derived from the Bayes
conditional update rule, where indeed no transition took place. This
behavior is evident from the spikes in Fig \ref{Reick-fig12}.

One would suspect that the additional information provided by
subdividing data into counts registered in the first and second half
of every time bin would only serve to yield a better estimate of the
atomic state, since no knowledge is lost by this finer binning of
the data. In the case where no transitions occur and we aim to
detect the state initially occupied by the atom in a quantum
nondemolition manner, the Bayesian analysis indeed becomes
independent of data binning size for a Poissonian count process. To
study this issue in our system, with the state-changing rate process
occurring simultaneously with the probing, we evaluated the one-atom
telegraph signals for different bin sizes using the Bayesian
algorithm, see fig. \ref{Reick-fig12}. Even for $\Delta t_b =$ 10
\textmu s, when there is no click in 80\% of all bins, the
calculated probability $p_0(t)$ is often close to 0 or 1, although
the state probability shows more short spikes compared to $\Delta
t_b =$ 1 ms.

\begin{figure}
\centering
\includegraphics[width=\columnwidth]{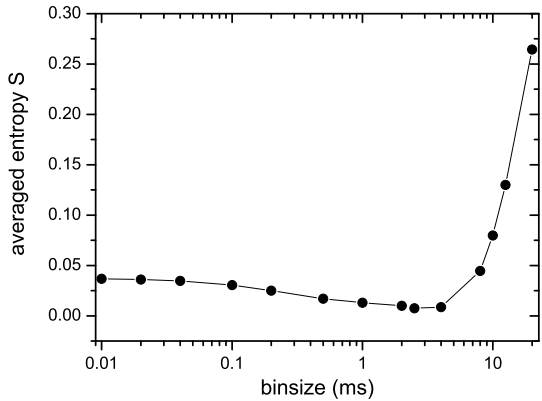}
\caption{Time- and ensemble averaged entropy $S$ as a function of
binning time $\Delta t_b$.}\label{Reick-fig13}
\end{figure}

To give a single quantitative measure of our uncertainty about the
atomic state, we calculate the entropy
\begin{equation} \label{eq:entropy}
S = \langle -\sum_{\alpha} p_{\alpha} \log p_{\alpha} \rangle,
\end{equation}
where the average $\langle...\rangle$ is performed over the whole
duration of all analyzed traces. The entropy is plotted in fig.
\ref{Reick-fig13} for a range of bin times from 10 \textmu s to 20
ms. The sharp rise of $S$ for large bins is due to the high
probability in every time bin for an atomic transition to occur. We
ascribe the increase of $S$ towards shorter bins to the occurrence
of more spikes in $p_{\alpha}(t)$, already visible in fig.
\ref{Reick-fig12} (d) and (e). According to the entropy measure,
there seems to be an optimum time bin, which is related to the
magnitude of the quantum jump rates. We recall, however, that the
entropy (\ref{eq:entropy}) is only one of many possible measures of
the information extracted from the system. If, for example, the
measurements constitute a component in a feedback mechanism, access
to data on the shortest time scale may yield the better performance
with respect to the desired goal of the feedback protocol.

\subsection{Origin and modeling of super-Poissonian count distributions}

The existence of an optimum bin time, leading to a minimum in the
time averaged entropy (\ref{eq:entropy}), is observed both for our
experimental histogram data and in simulations with Poissonian
counting statistics associated with each atomic state. The case of
super-Poissonian counting distributions, i.e., distributions with a
variance exceeding the mean value of the number of counts, however,
presents it own separate problems, and points to more elaborate
future methods of analysis.

We already commented on the apparent extra fluctuations in the light
transmission signal being possibly correlated with the atomic motion
between sites exhibiting different coupling strengths to the cavity
mode, corresponding to different transmission levels. This suggests
an extended model, where the state with no atoms coupled to the
field ($\alpha=0$) is retained as a single state, while states with
$\alpha=1,2$ are split according to an extra position label,
attaining a number of different values. If for example a single atom
can reside in two locations leading to two different Poissonian
transmission signals, the long time averaged photon count
distribution will be a weighted sum of these distributions, while
the count number correlation function within an experimental trace
may reveal the transition rates between the atomic locations,
equivalent to our analysis of internal state transitions in sec.
\ref{subsec:OneAtomBayes}. This is an appealing and very likely
explanation of the broadened histograms, and it points to an
interesting problem for our previous analysis.

If the super-Poissonian fluctuations in our counting histograms are
caused by atomic motion between states with different Poissonian
signals, counts in close lying time bins, where atoms have not yet
moved, should be correlated. This implies that the Bayesian update
is no longer a Markov process, where the updated probabilities
depend only on the most recent value and the latest measurement
result, but also knowledge of previous counts should be applied to
extract maximum information about the atomic state. This effect may
have significant consequences for very short time bins, where each
bin offers a low signal-to-noise ratio, but where correlations
between bins may be strong. We have analyzed our experimental
records, and we indeed find such correlations, but because of
limited statistics these findings could not be incorporated
quantitatively into our analysis. This does not imply that our
previous use of the Bayes update formalism produces erroneous
results, but it should be noted that it represents the update based
on a restricted access to (or memory of) the measurement data, and
hence it provides a non-optimal estimate of the atomic state based
on incomplete information.

\section{Outlook}

We have shown that a Bayesian analysis of experimental transmission
signals from a cavity containing one or two atoms provides a high
degree of certainty about the atomic state. We have demonstrated how
rates of the atomic processes can be fitted to the data, and we have
discussed possible physical explanations of noise in the data beyond
the predictions of simple models.

A natural next step would be to use a more complete model, including
the larger number of position states and internal states of the
atoms. We recall that already for the simplest model with only
internal state dynamics, finding the parameters is not a trivial
task, but ad hoc iterative procedures have allowed the
identification of consistent sets of parameters used in our present
analysis in this paper.

It will put stringent demands on the reproducibility of large data
sets to make a reliable fit to more advanced models, but we wish to
conclude this paper with a brief mentioning of a promising
systematic theoretical data analysis that can be applied to such
data in a future more elaborate treatment: the Hidden Markov Model
(HMM) \cite{Cappe2000}. We indicated that there is a possible
physical mechanism responsible for the fluctuations and for the
temporal correlations between count signals. In this way we point at
an underlying Markovian model, where the atoms perform transitions
between different internal and position states, and for each of
these states, the coherent light field is transmitted with a
definite transmission coefficient, and counting statistics are
Poissonian with no temporal correlations. This is, indeed, a
physical realization of a Hidden Markov Process in statistical
modelling of time series, with applications in insurance, finance,
speech recognition, image analysis and many other fields, where a
single series of data is mathematically modeled as the outcome of a
system undergoing transitions between (hidden) states, each giving
different data characteristics. In their most advanced forms, HMMs
only assume the transitions between the hidden states to be
Markovian, i.e., the state populations follow a transfer matrix of
discrete or continuous population changes, while the signal can have
any state dependent probability distribution.

Our problem belongs to a narrower class with continuous rate
equations (with unknown rates), and it is plausible to assume
Poissonian count statistics parameterized by a single parameter for
each atomic state. This case is treated, e.g., in \cite{Paroli2000},
and the problem of estimating the transition rates among a family of
$N$ states and the $N$ photon transmission rates from the data is
solved by an iterative variational application of the maximum
likelihood principle. In a genuine HMM, the number of states $N$ is
not known, and one merely attempts to fit the data with different
candidate numbers of states. For an application to our problem, we
are guided by the physics, and after a successful fit, we would
request that the states identified should have the properties
corresponding to a few position states for each of the internal
state $\alpha=1,2$ cases. I.e., they should occur in groups with
similar photon scattering rates, and certain transition rates should
be very small or vanish.

In addition to an extended model for the analysis, we aim at
improving the experimental conditions, such that the
super-Poissonian noise is less pronounced. Since we attribute these
fluctuations mainly to thermal motion of the atom, increasing the
stability of the coupling strength requires a tighter confinement of
the atom. This could be achieved by employing cavity-mediated
cooling forces \cite{Domokos2003,Murr2006}, Raman cooling
\cite{Boozer2006} or additional trapping potentials.

The rate at which information about the atom-cavity system can be
acquired is ultimately limited by the photon flux arriving at the
detector. The most important obstacles for further enhancement of
the detection efficiency are losses from the cavity-mirror coatings
and the limited quantum efficiency of the SPCM. Employing homodyne
or heterodyne detection would permit the use of detectors with a
quantum efficiency close to 100\%. The former problem could be
solved by using a more open cavity configuration, where the
transmission coefficient is significantly larger than the losses.

Advancements both in terms of experimental conditions and
statistical analysis could finally lead to the development and
implementation of quantum feedback techniques for the preparation,
stabilization and error correction of non-classical quantum states
\cite{Carvalho2007,Carvalho2008}.

\section*{Acknowledgements}

We acknowledge financial support by the EC (IP SCALA). S. R.
acknowledges support from the ``Deutsche Telekom Stiftung'' and T.
K. acknowledges support from the ``Studienstiftung des Deutschen
Volkes''.


\begin{thebibliography}{10}
\bibitem{footnote1}
The $m_F$-distribution within the $\ket{F=4}$ manifold for a
coordinate system
  where the quantisation axis is parallel to the electric field, i.e. for
  $\pi$-transititions, is calculated to be 34.4\% in $m_F = 0$, 23.9 \% in
  $m_F=\pm1$, 7.8 \% in $m_F=\pm 2$, 1.1\% in $m_F=\pm 1$, and 0.1\% in
  $m_F=\pm4$.
\end{thebibliography}
\end{document}